 \documentclass[twocolumn,amsmath,showpacs,amsfonts,aps,prc,floatfix]{revtex4}
\usepackage{graphicx}
\usepackage{bm}

\begin{document}

\title{Direct photon production and interferometry in $\sqrt{s}_{NN}$=200 GeV Au+Au and in
$\sqrt{s}_{NN}$=2.76 TeV Pb+Pb collisions}
 
\author{A. K. Chaudhuri}
\email[E-mail:]{akc@vecc.gov.in}
\affiliation{Variable Energy Cyclotron Centre,\\ 1/AF, Bidhan Nagar,
Kolkata 700~064, India}

\begin{abstract}

Direct photon production, in  $\sqrt{s_{NN}}$=200 GeV Au+Au and in $\sqrt{s_{NN}}$=2.76 TeV Pb+Pb    collisions, are studied in a hydrodynamic model.
 Ideal hydrodynamic model, initialised to reproduce experimentally charged particle spectra in RHIC and LHC energy, also reproduces the PHENIX and ALICE measurements for the direct photon spectra in Au+Au collisions at RHIC and Pb+Pb collisions at LHC. The model however produces less elliptic flow than in experiment. 
Discrepancy between experiment and hydrodynamic simulation is comparatively less in Pb+Pb collisions at LHC than in Au+Au collisions at RHIC. We also studied direct photon correlation and determined the HBT radii. In  0-10\%-50-60\% collisions, HBT radii in Au+Au or in Pb+Pb  collisions   do not show large centrality dependence. Energy dependence of the HBT radii for direct photons is also not large.
  \end{abstract} 


\maketitle
\section{Introduction}

Recent experiments in $\sqrt{s_{NN}}$=200 GeV Au+Au collisions   at Relativistic Heavy Ion Collider (RHIC)  \cite{BRAHMSwhitepaper}\cite{PHOBOSwhitepaper}\cite{PHENIXwhitepaper}\cite{STARwhitepaper} and    $\sqrt{s_{NN}}$=2.76 TeV Pb+Pb collisions at Large Hadron Collider (LHC) \cite{Aamodt:2010pb}\cite{Collaboration:2010cz}\cite{Aamodt:2010jd}\cite{Aamodt:2010pa}\cite{:2011vk}, produced convincing evidences that in Au+Au collisions at 
RHIC and in Pb+Pb collisions at LHC, a deconfined medium or Quark-Gluon Plasma (QGP) is produced.
Strong elliptic flow in non-central collisions is the key evidence to this
understanding. Hydrodynamical analysis of experimental charged particles data , also suggests that in Au+Au and in Pb+Pb  collisions, a collective medium, with viscosity to entropy ratio close to the AdS/CFT lower bound of viscosity,   $\eta/s \geq 1/4\pi$ is produced \cite{QGP3}\cite{Chaudhuri:2009uk}\cite{Chaudhuri:2008ed}\cite{Roy:2011xt}\cite{Chaudhuri:2011hv}.  
 
Direct photons probe  the {\em early} medium produced in a collision better than the charged hadrons. 
Hadrons, being strongly interacting, are emitted from the  surface of the thermalised matter and
carry information about the freeze-out surface only. They are unaware of
the condition of the interior of the matter and can provide information about the deep interior only in an indirect way.
In a hydrodynamic model, one fixes the initial conditions of the fluid  such that the "experimental" freeze-out surface is correctly reproduced.  
In contrast 
to hadrons,    photons, being weakly interacting, are emitted from 
whole volume of the matter. Throughout the evolution of the matter, photons are emitted. Conditions of the produced matter, at its deep interior,
are better probed by the photons. Depending on the transverse momentum, direct photons can
probe different aspects of heavy ion collisions. A thermalised
medium of quarks and gluons, or of hadrons, can produce
significant number of thermal photons. They are
low $p_T$ photons ($p_T \leq 3$ GeV/c). Low $p_T$ photons can test whether or not, QGP is produced in the collisions. Hard  
photons ($p_T >$ 6 GeV/c) are of pQCD origin and test the
pQCD models. Fast partons from 'jet' can interact with 
thermal partons of QGP and produce photons. 
At intermediate $p_T$ range, ($3\leq p_T \leq$ 6 GeV/c), interaction jets with QGP could be an important source of direct photons \cite{Fries:2002kt}\cite{Gale:2005zd}.

Recently, ALICE collaboration measured direct photon production in 2.76 TeV Pb+Pb collision at LHC. \cite{Lohner:2012ct}. $p_T$ spectra as well as elliptic flow in 0-40\% collision was measured. Inverse slope parameter of an exponential fit to the low $p_T$ part of the spectra give a temperature $T=304\pm 51$ MeV. Previously, PHENIX collaboration
measured direct photon $p_T$ spectra and elliptic flow in $\sqrt{s}_{NN}$=200 GeV Au+Au collisions  \cite{Afanasiev:2012dg}\cite{Adare:2011zr}  \cite{Miki:2008zz}\cite{Adler:2005ig}. In 0-20\% Au+Au collisions, inverse slope parameter is only $T=220\pm 19$ MeV, considerably lower than the slope parameter at LHC energy. In ideal hydrodynamic models, 
photon production in Au+Au collisions at RHIC energy has been studied extensively \cite{Fries:2002kt} \cite{Alam:2007dv}\cite{Chatterjee:2005de}\cite{Gale:2008wf}  \cite{Liu:2009kta} \cite{Liu:2009kq}\cite{Liu:2010rd}\cite{Helenius:2013bya} \cite{vanHees:2011vb}\cite{Holopainen:2011pd}. Recently, it was shown that direct photon production in $\sqrt{s}_{NN}$=2.76 TeV Pb+Pb collisions is also reproduced in a hydrodynamic model  \cite{Nayak:2012dj}. 
In the present paper, in ideal hydrodynamic model, we have analysed the 
direct photon spectra and elliptic flow in Pb+Pb collisions at LHC as well as in Au+Au collisions at RHIC. In some earlier publications \cite{Chaudhuri:2009uk}\cite{Chaudhuri:2008ed}\cite{Roy:2011xt}\cite{Chaudhuri:2011hv}, in hydrodynamical model, we have analysed the charged particles $p_T$ spectra and elliptic flow in $\sqrt{s}_{NN}$=200 GeV Au+Au and in $\sqrt{s}_{NN}$=2.76 TeV Pb+Pb collisions. To analyse the direct photon spectra in Au+Au and in Pb+Pb collisions, we initialise the fluid with similar initial conditions. It may be mentioned here that, though ideal hydrodynamic model explains a large part of the experimental data on charged particles,  the best description to the data is obtained in dissipative hydrodynamics,  with viscosity over entropy ratio close to the AdS/CFT conjectured value \cite{Policastro:2001yc}\cite{Kovtun:2003wp}, $\eta/s \geq 1/4\pi$.

In the following, results of our simulations for direct photons in ideal hydrodynamic model will be discussed in details. Briefly, both in $\sqrt{s}_{NN}$=200 GeV Au+Au collisions and in $\sqrt{s}_{NN}$=2.76 TeV Pb+Pb collisions, ideal hydrodynamical model simulations for direct photons, well reproduces the experimentally measured spectra.  The elliptic flow data are however, underpredicted, more in Au+Au collisions than in Pb+Pb collisions. We have also simulated HBT correlation for two photons and extracted HBT radii, as a function of centrality. Even though, from RHIC to LHC, collision energy increases by a factor of $\sim$14, HBT radii for direct photons do not show large energy dependence. HBT radii for direct photons  also do not show large centrality dependence.

\section{Hydrodynamical model for direct photon production}

In a hydrodynamical model, the invariant distribution of direct photons is obtained by convoluting the  photon production rate with space-time evolution of the fluid. 
Invariant distribution of direct photons can be obtained as,

\begin{equation}
E\frac{dN^\gamma}{d^3p_T}=\int d^4x \left ( E\frac{dR^\gamma}{d^3p_T} \right ) 
\end{equation} 
 
\noindent where  $\left ( E\frac{dR^\gamma}{d^3p_T} \right ) $ is the direct photon production rate and $d^4x$ is the elementary fluid volume.

Several authors have studied photon emission rate hot hadronic gas \cite{Kapusta:1991qp}\cite{Nadeau:1992cn}\cite{Turbide:2003si}. In the following we use the parametrised form given in \cite{Turbide:2003si}. Following processes are included,

(a) $\pi+ \rho \rightarrow \pi + \gamma$,
(b) $\pi + \pi \rightarrow \rho + \gamma$, 
(c) $\rho \rightarrow \pi + \pi + \gamma$,
(d) $\pi+K^*\rightarrow K + \gamma$,
(e) $\pi+ K\rightarrow K^* + \gamma$,
(f) $\rho + K \rightarrow K + \gamma$,
(g) $K^* + K \rightarrow \pi + \gamma$,

  Direct photon emission rate from QGP is also well studied \cite{Kapusta:1991qp}\cite{Aurenche:1998nw}\cite{Aurenche:1999tq}\cite{Aurenche:2000gf}\cite{Arnold:2001ms}. In \cite{Arnold:2001ms}, photon emission rates
from QGP are calculated in the leding order. We use the parametrised form given there.
 
As it was mentioned earlier, invariant photon yield is obtained by convoluting the elementary production rates over space-time evolution. Assuming that in Au+Au collisions at RHIC or in Pb+Pb collisions at LHC, a baryon free QGP fluid is produced, the
space-time evolution of the fluid is obtained by solving the    energy-momentum conservation equation, 
 
\begin{eqnarray}  
\partial_\mu T^{\mu\nu} & = & 0,\label{eq1}\\    
T^{\mu\nu}&=&(\varepsilon+p)u^\mu u^\nu - pg^{\mu\nu}, \label{eq2} .
\end{eqnarray}

$\varepsilon$, $p$ and $u$ in Eq.\ref{eq2} are the energy density, pressure and fluid velocity respectively.  
Assuming boost-invariance, Eqs.\ref{eq1}  are solved in $(\tau=\sqrt{t^2-z^2},x,y,\eta_s=\frac{1}{2}\ln\frac{t+z}{t-z})$ coordinates, with the code   "`AZHYDRO-KOLKATA"', developed at the Cyclotron Centre, Kolkata.
 Details of the code can be found in \cite{Chaudhuri:2008sj}.
 
 \begin{table}[h] 
\caption{\label{table1} Central energy density and temperature of fluid for hydrodynamical analysis for $\sqrt{s}_{NN}$=200 GeV Au+Au and $\sqrt{s}_{NN}$=2.76 TeV Pb+Pb collisions, at the initial time $\tau_i$=0.6 fm. Initial transverse fluid velocity is assumed zero, $v_x=v_y=0$.}
 \begin{ruledtabular} 
  \begin{tabular}{|c|c|c|c|}\hline
System & $\sqrt{s}_{NN}$ & $\epsilon_0$ ($GeV/fm^3$) & T(MeV) \\ \hline 
Au+Au & 200 & 30.0 &370 \\
Pb+Pb & 2760 &125 &530\\
\end{tabular}\end{ruledtabular}  
\end{table} 

Hydrodynamic equations (Eqs.\ref{eq1}) are closed with an equation of state $p=p(\varepsilon)$.
 Currently, there is consensus that the confinement-deconfinement transition is a cross over. The cross over or the pseudo critical temperature for the quark-hadron transition  is
$T_c\approx$170 MeV \cite{Aoki:2006we,Aoki:2009sc,Borsanyi:2010cj,Fodor:2010zz}.
In the present study, we use an equation of state where the Wuppertal-Budapest \cite{Aoki:2006we,Borsanyi:2010cj} 
lattice simulations for the deconfined phase is smoothly joined at $T=T_c=174$ MeV, with hadronic resonance gas EoS comprising of all the resonances below mass $m_{res}$=2.5 GeV. Details of the EoS can be found in \cite{Roy:2011xt}.

 \begin{figure}[t]
 \vspace{1cm}
\center
\resizebox{0.35\textwidth}{!}{%
  \includegraphics{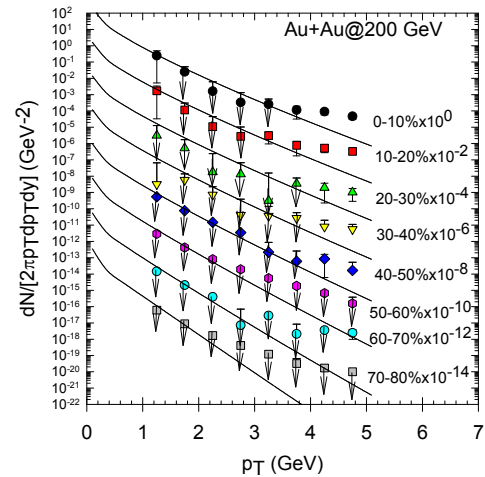}
}
\caption{(color online) The colored symbols are PHENIX measurements for 
transverse momentum spectra for photons in $\sqrt{s}_{NN}$=200 GeV Au+Au collisions.
The lines are predictions from hydrodynamic model. 
 }\label{F1}
\end{figure}

Solution of partial differential equations (Eqs.\ref{eq1},\ref{eq2}) requires initial conditions, e.g.  transverse profile of the energy density ($\varepsilon(x,y)$), fluid velocity ($v_x(x,y),v_y(x,y)$)   at the initial time $\tau_i$. A freeze-out temperature is also needed. Charged particles production in $\sqrt{s}_{NN}$=200 GeV Au+Au and in $\sqrt{s}_{NN}$2.76 TeV Pb+Pb collisions have been analysed in hydrodynamic models \cite{Chaudhuri:2009uk}\cite{Chaudhuri:2008ed}\cite{Roy:2011xt}\cite{Chaudhuri:2011hv}.
Transverse energy density profile can be obtained in a Glauber model. At impact parameter $b$, the initial energy density is distributed as,

\begin{equation}\label{eq4}
\varepsilon({\bf b},x,y)=\epsilon_0 [(1-f)N_{part}({\bf b},x,y)+fN_{coll}({\bf b},x,y)]
\end{equation}

$N_{part}$ and $N_{coll}$ in Eq.\ref{eq4} are the transverse profile of
the average number of participants and average number of binary
collisions respectively. They are calculated in the Glauber model. $f$ is the hard scattering fraction.
We assume that in $\sqrt{s}_{NN}$=2.76 TeV Pb+Pb collisions are dominated by hard scattering and $f= 0.9$ \cite{Roy:2011xt}\cite{Chaudhuri:2011hv} . In $\sqrt{s}_{NN}$=200 GeV Au+Au collisions, hard scattering fraction is assumed to be 25\% \cite{Chaudhuri:2009uk}\cite{Chaudhuri:2008ed}.

With the central energy density and temperature of the fluid,  listed in table.\ref{F1}, several experimental data, e.g. centrality dependence of charged particles multiplicity, $p_T$ spectra, elliptic flow etc. in Au+Au collisions at RHIC and Pb+Pb collisions at LHC are reasonably well explained.
For simulating photon production, we then use the parameters listed in table.\ref{table1}. Initial transverse fluid velocity is assumed zero. The freeze-out temperature is assumed to be $T_F$=110 MeV. 

\section{Direct photons in $\sqrt{s}_{NN}$=200 GeV Au+Au
and $\sqrt{s}_{NN}$=2.76 TeV Pb+Pb collisions}

\subsection{Transverse momentum spectra}

PHENIX collaboration studied direct photon production in $\sqrt{s}_{NN}$=200 GeV Au+Au collisions \cite{Afanasiev:2012dg}. PHENIX measurements for the transverse momentum distribution of direct photons in 0-10\%-70-80\% collision centralities are shown in  Fig.\ref{F1}. The solid lines in Fig.\ref{F1} are hydrodynamical model simulations. Ideal hydrodynamic model reproduces the spectra reasonably well. In central and mid-central  collisions (0-10\%-40-50\%), experimental data upto $p_T\approx$3 GeV are reproduced. Data   are however underpredicted at larger $p_T$. It is understood. As it was discussed earlier, at larger $p_T$ photons, apart from the thermal photons, there are other sources, e.g. pQCD photons, jet photon etc.  In more peripheral collisions, ideal hydrodynamic simulation reproduces experimental data in much more limited $p_T$ range.
The result is not unexpected. In peripheral collisions, experimental data possibly demand dissipative effects like shear viscosity, and ideal hydrodynamic model fails to reproduce experimental charged particles spectra. 
Also, in peripheral collisions non-thermal photon production becomes more important and  spectra with only thermal production then deviate from the data.

  \begin{figure}[t]
\center
\resizebox{0.35\textwidth}{!}{%
  \includegraphics{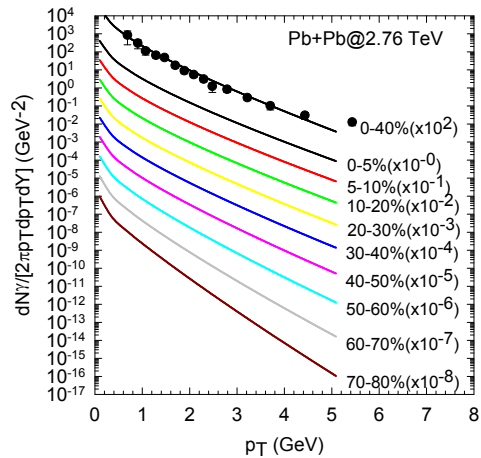}
}
\caption{(color online) The filled circles are  ALICE measurements for 
transverse momentum spectra for photons in $\sqrt{s}_{NN}$=2.76 TeV Pb+Pb collisions. 
The line is the prediction from hydrodynamic model. 
 }\label{F2}
\end{figure}  

Recently ALICE collaboration published their measurements of direct photon \cite{Lohner:2012ct}. Transverse
a momentum spectrum of direct photons in 0-40\% Pb+Pb collisions is shown in Fig.\ref{F2}. 0-40\% collisions approximately corresponds to impact parameter b=8.9 fm Pb+Pb collisions.
The solid line in Fig.\ref{F2} is the hydrodynamic simulation for photon spectra in  b=6.9 fm Pb+Pb collisions. Direct photon yield upto $p_T$=4 GeV are correctly reproduced. Good description of direct photon production in $\sqrt{s}_{NN}$=200 GeV Au+Au and $\sqrt{s}_{NN}$=2.76 TeV Pb+Pb collisions gave us confidence to predict for the photon yield as a function of collision centrality. In Fig.\ref{F2}, hydrodynamic predictions for direct photons in 0-10\%-70-80\% in $\sqrt{s}_{NN}$=2.76 TeV Pb+Pb collisions are shown. Considering that in peripheral Au+Au collisions, hydrodynamic predictions underestimate photon production at large $p_T$,  the predictions given here may also underestimate photon production at $p_T >$ 3 GeV.  As expected, in more peripheral collisions,
inverse slope of the spectra decreases, indicating less fluid temperature.  In several experiments, slope of the direct photon spectra is used to estimate the fluid temperature.
In table.\ref{table2}, the inverse slope of the simulated spectra in the $p_T$ range 1-2 GeV, 
in $\sqrt{s}_{NN}$=200 GeV Au+Au and in $\sqrt{s}_{NN}$=2.76 TeV Pb+Pb
collisions, are listed as a function of collision centralities. Compared to 200 GeV Au+Au collisions, inverse slope is always greater in 2.76 TeV Pb+Pb collisions. For comparison, 
in table.\ref{table2}, I have listed the initial   central fluid temperature $T_0$.
Central fluid temperature is much larger than the inverse slope parameter. It is expected also, temperature obtained from spectra are in a sense averaged over
evolution period and is smaller than the initial central temperature. Thus fluid can be  produced at much higher temperature than estimated from the photon spectra.

 \begin{table}[h] 
\caption{\label{table2} Centrality dependence of inverse slope parameter obtained from
exponential fit to direct photon spectra in   $\sqrt{s}_{NN}$=200 GeV Au+Au and $\sqrt{s}_{NN}$=2.76 TeV Pb+Pb collisions are listed. Also listed is the initial central  fluid temperature ($T_0$).}
 \begin{ruledtabular} 
  \begin{tabular}{|c|c|c|c|c||}\hline
 & \multicolumn{2}{c|}{Au+Au@200GeV}&\multicolumn{2}{c|}{Pb+Pb@2.76TeV}\\ \hline
centrality & Inv.slope & $T_0$& Inv.slope&$ T_0$\\ 
  &(MeV) &(MeV)&(MeV) &(MeV) \\ \hline\hline 
 0-10 & 219& 368 &260 &523\\
 10-20& 217& 354&258& 505\\
20-30& 214&  338&254& 484\\ 
30-40& 210&  320&249& 458\\ 
40-50& 206&  296&240& 424\\
50-60& 198&  267&228& 383\\
60-70& 191&  236&215& 333\\ 
70-80& 185&  203&202& 281\\  
\end{tabular}\end{ruledtabular}  
\end{table}

 \begin{figure}[t]
\center
\vspace{1cm}
\resizebox{0.35\textwidth}{!}{%
  \includegraphics{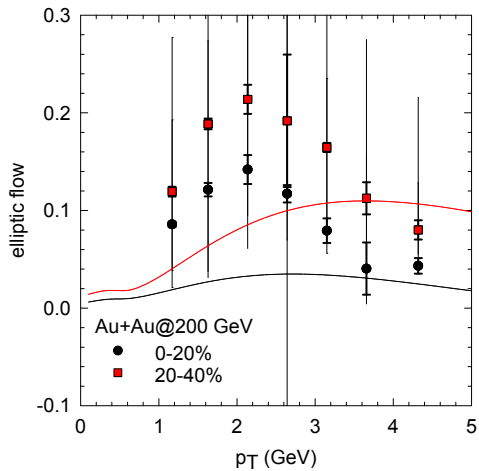}
}
\caption{(color online) PHENIX measurements direct photon  
elliptic flow  in 0-20\% and 20-40\%  Au+Au collisions at RHIC are shown. Both the statistical and systematic errors are shown by the thick and thin error bars. The black   and red lines are hydrodynamic model hydrodynamical model simulations for the same. 
 }\label{F3}
\end{figure}
  
 \begin{figure}[t]
\center
\resizebox{0.35\textwidth}{!}{%
  \includegraphics{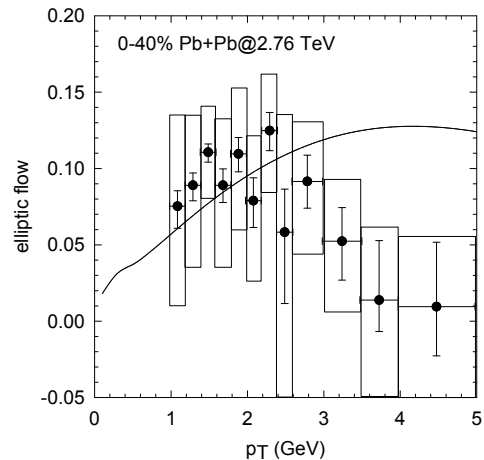}
}
\caption{The filled circles are  ALICE measurements for direct photon elliptic flow   in
0-40\%  Pb+Pb collisions at LHC. The statistical and systematic erros are shown.
The line is the prediction from hydrodynamic model. 
 }\label{F4}
\end{figure}  

 \begin{figure}[t]
\center
\vspace{1cm}
\resizebox{0.35\textwidth}{!}{%
  \includegraphics{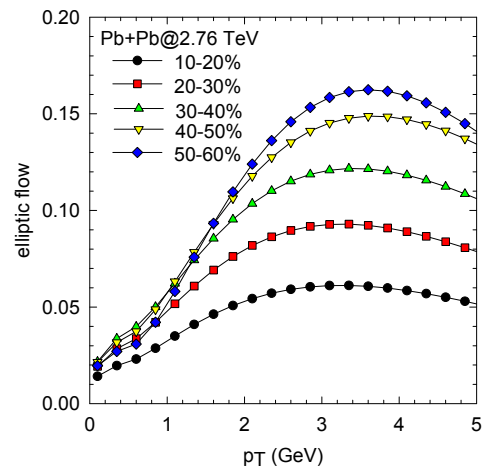}
}
\caption{(color online) Simulated elliptic flow in $\sqrt{s}_{NN}$=2.76 TeV Pb+Pb collisions.
Results for 10-20\% -50-60\% collision centralities are shown. }\label{F5}
\end{figure}  

\subsection{Elliptic flow}
 
Elliptic flow is one of the most important observables in RHIC and LHC energy collisions.
Since photons are continuously emitted from the fireball, elliptic flows of photon are more powerful than that of hadrons, which are emitted only from the freeze-out surface.
PHENIX measurements  \cite{Adare:2011zr} for elliptic flow in 0-20\% and 20-40\% Au+Au collisions are shown in Fig.\ref{F3}. At low $p_T$, $v_2$ increases, reaching a maxima around $p_T\sim 2 GeV$, decreases thereafter. The solid lines in Fig.\ref{F3} are the hydrodynamic model simulations. In Au+Au collisions, hydrodynamic model largely underpredict the data. The results are not 
unexpected. In an earlier simulation \cite{Chatterjee:2005de}, direct photon elliptic flow was simulated in Au+Au collisions. It was shown that even in very peripheral collision, hydrodynamical evolution generate very little flow.   Recently, ALICE collaboration \cite{Aamodt:2010pa} published direct photon elliptic flow in 0-40\% Pb+Pb collisions at $\sqrt{s}_{NN}$=2.76 TeV. The measurements are shown in Fig.\ref{F4}. Experimental data has large error bars, still, qualitative behavior of direct photon elliptic flow in Pb+Pb collisions at LHC is similar to that in Au+Au collisions at RHIC.
At low $p_T$ $v_2$ increase, reaching a maxima around $p_T\sim 2 GeV$ and then decreases at higher $p_T$. In Fig.\ref{F4}, hydrodynamic model simulation for $v_2$ is shown as the 
solid line.  At low $p_T < 2 GeV$, hydrodynamic simulation marginally underestimate the experimental flow. Discrepancy between hydrodynamic model simulations and experimental  elliptic flow is considerably less than that in Au+Au collisions at RHIC. At high $p_T$ however, model predict more flow than in experiment.
In Fig.\ref{F5}, hydrodynamic model simulations for direct photon flow in 10-20\%, 20-30\%, 30-40\%, 40-50\% and 50-60\% Pb+Pb collisions at LHC are shown. They are shown here as a reference only. We expect that at low $p_T$, $p_T \leq$2 GeV, experimentally measured flows will be close to the simulated flows.

\section{Photon interfereometry}

Hanbury-Brown-Twiss (HBT) interferometry is widely used in the analysis of heavy ion collision data to estimate the size of the particle emitting source. Spin averaged intensity correlation between two photons of momenta $\bf{k}_1$ and $\bf{k}_2$, from a chaotic source can be expressed as \cite{Wiedemann:1999qn}\cite{Frodermann:2009nx}\cite{Srivastava:1993pt}  \cite{Alam:2003gx},

\begin{eqnarray}
C({\bf q},{ \bf K})&=&1+\frac{1}{2}\frac{ \left | {  \int d^4x S(x,\bf{K}) e^{i.q.x}}\right |^2}
{\int d^4x S(x,\bf{k_1}) \int d^4x S(x,\bf{k_2})} \nonumber \\
&\approx & 1+\frac{1}{2}\frac{ \left | {  \int d^4x S(x,\bf{K}) e^{i.q.x}}\right |^2}
{ \left | \int d^4x S(x,\bf{K}) \right |^2} \label{eq4}
\end{eqnarray}

\noindent where $\bf{q}=\bf{k_1}-\bf{k_2}$ is the relative momentum and $\bf{K}=\frac{\bf{k_1}+\bf{k_2}}{2}$  is the average momentum. 
The emission function $S(x,\bf{K})$ is the probability of emitting a photon from the source point $x$ with momentum ${\bf K}$. It is generally approximated by the rate of production of photon.
The last equation is obtained with the 'smoothing approximation' that the emission function $S(x,{\bf k})$ varies  slowly
over the momentum range where the correlator deviates
from unity, $S(x,{\bf k_1})\approx S(x,{\bf k_2})=S(x,{\bf K})$. 

The correlation function $C(\bf{q},\bf{K})$ is generally decomposed in terms of the outward, sideward, and longitudinal momentum differences;
$q_{out}$, $q_{side}$ and $q_{long}$ respectively. One can write the
four-momentum of the i-th photon in-terms of
transverse momentum $k_T$ , rapidity y, and azimuthal
angle $\psi$ as,

\begin{equation}
k^\mu_i=(k_{iT}\cosh y_i, k_{iT}\cos \psi_i, k_{iT} \sin \psi_i, k_{iT} \sinh y_i)
\end{equation}

 \begin{figure}[t]
\center
\resizebox{0.45\textwidth}{!}{%
  \includegraphics{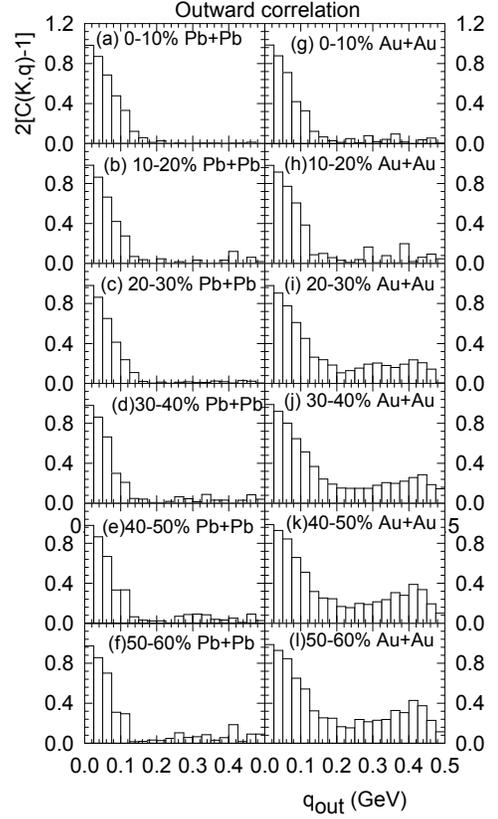}
}
\caption{Hydrodynamical simulation for outward correlation for direct photons, in $\sqrt{s}_{NN}$=2.76 TeV Pb+Pb and in $\sqrt{s}_{NN}$=200 GeV Au+Au collisions, in 0-10\% -50-60\% collision centrality. 
 }\label{F6}
\end{figure}      

The three momentum differences are,

\begin{eqnarray}
q_{out}&=&\frac{\bf{q_T}\cdot \bf{K_T} }{ K_T} \nonumber \\
&=& \frac{(k_{1T}^2-k_{2T}^2)} {\sqrt{k_{1T}^2+k_{2T}^2+2k_{1T}k_{2T}cos(\psi_1-\psi_2) } }\\
q_{side}&=& {\bf q_T}- q_0 \frac{ {\bf K_T}}{ K_T} \nonumber\\
&=&\frac{2k_{1T}k_{2T} \sqrt{1-cos^2(\psi_1-\psi_2)} } {\sqrt{k_{1T}^2+k_{2T}^2+2k_{1T}k_{2T}cos(\psi_1-\psi_2) } }\\
q_{long}&=&k_{1z}-k_{2z} \nonumber\\
&=&k_{1T}\sinh y_1-k_{2T}\sinh y_2
\end{eqnarray}

The main characteristic of the correlation function is
the q-range over which it decays to 1. The corresponding
half-widths of the correlator are called HBT radius parameters
or HBT radii. In the absence of final state interactions, HBT radii
are generally  extracted by approximating the correlation function by a Gaussian source,

\begin{eqnarray}
C(q_{long},{\bf K})&=&1+\frac{1}{2} e^{\left [-q_{out}^2 R_{out}^2-q_{side}^2 R_{side}^2-q_{long}^2 R_{long}^2 \right ]}, \nonumber \\
\end{eqnarray}

\noindent where, $R_{out}$, $R_{side}$ and $R_{long}$ are respectively called, outward radius, sideward radius and longitudinal radius.
We would like to note here that the
HBT radii $R_{out}$, $R_{side}$ or $R_{long}$, do not give directly the size of the emission region, rather they
  corresponds to the homogeneous region from which it is most likely that photon pair with momentum $K$ is emitted \cite{Wiedemann:1999qn}. It is generally a small fraction of the total fireball size. Furthermore, HBT radii only measure certain combination of spatial and temporal width. Within some model approximations, $R_{out}$ and $R_{long}$ measure the emission size and duration of particle emission, $R_{side}$ measures the transverse size \cite{Wiedemann:1999qn}. The ratio $R_{out}/R_{side}$ then indicate the duration of the particle emission.  
  





  \begin{figure}[t]
\center
\resizebox{0.45\textwidth}{!}{%
  \includegraphics{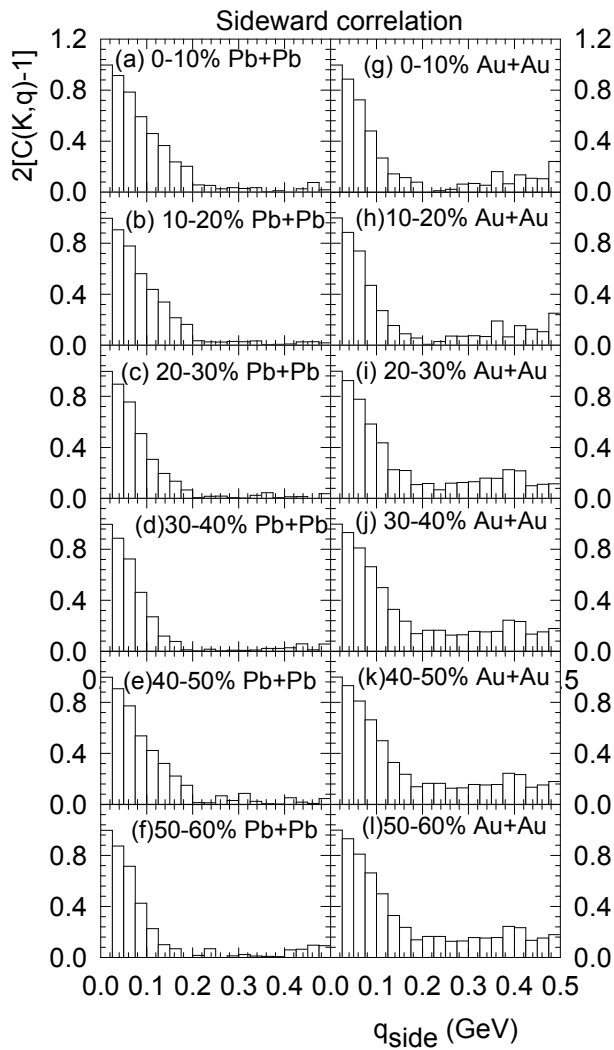}
}
\vspace{0.5cm}
\caption{Hydrodynamical simulation for sideward correlation function for direct photons in $\sqrt{s}_{NN}$=2.76 TeV Pb+Pb and in $\sqrt{s}_{NN}$=200 GeV Au+Au collisions, in 0-10\% - 50-60\% collision centrality.   
 }\label{F7}
\end{figure}


We have solved hydrodynamic equations with the assumption of boost-invariance. In  a model with boost-invariance, one can not comment on the longitudinal correlation. In the following, we  concentrate on the outward and sideward correlations only. We use a Monte-Carlo sampling programme to obtain the outward, and sideward correlation functions  from Eq.\ref{eq4}, for the direct photons in the transverse momentum range $0 \leq k_T \leq 1$ GeV. 
In  the left panels (a)-(f) of Fig.\ref{F6}, hydrodynamic simulation for the outward correlation, for direct photons in 0-10\%, 10-20\%, 20-30\%, 30-40\%, 40-50\% and 50-60\% $\sqrt{s}_{NN}$=2.76 TeV Pb+Pb collisions are shown. The same in $\sqrt{s}_{NN}$=200 GeV Au+Au collisions are shown in the right panels (g)-(l). The outward correlation functions were obtained for, $0 \leq k_{T1} \leq 1 GeV$,$0 \leq k_{T2} \leq 1$, and imposing the condition $\psi_1=\psi_2=\psi$, with $0\leq \psi \leq 2\pi$, such that $q_{side}=0$. In $\sqrt{s}_{NN}$=2.76 TeV Pb+Pb collisions, the correlation functions approximately corresponds to Gaussian shape.
Only in peripheral collisions and at large momentum transfer $q_{out}$, the correlation function depart from the Gaussian shape, even then, the departure is small.
  In Au+Au collisions however, even in central and mid-central collisions, the correlation function deviate from the Gaussian shape. Width of the outward correlation function   in $\sqrt{s}_{NN}$=200 GeV Au+Au collisions is also more than that in $\sqrt{s}_{NN}$=2.76 TeV Pb+Pb collisions.  

In Fig.\ref{F7}, simulation results obtained for the sideward correlation functions are shown. Sideward correlation is obtained for,
$0\leq \psi_1 \leq 2\pi$, $0\leq \psi_2 \leq 2\pi$ and   imposing the condition that $k_{T1}=k_{T2}=k_T$, $0 \leq k_T \leq 1 GeV $, such that $q_{out}$ is identically zero. As it was for outward correlation, sideward correlation functions also depart more from the Gaussian shape in $\sqrt{s}_{NN}$=200 GeV Au+Au collisions than in $\sqrt{s}_{NN}$=2.76 TeV Pb+Pb collisions. 

\begin{figure}[t]
\center
\resizebox{0.35\textwidth}{!}{%
  \includegraphics{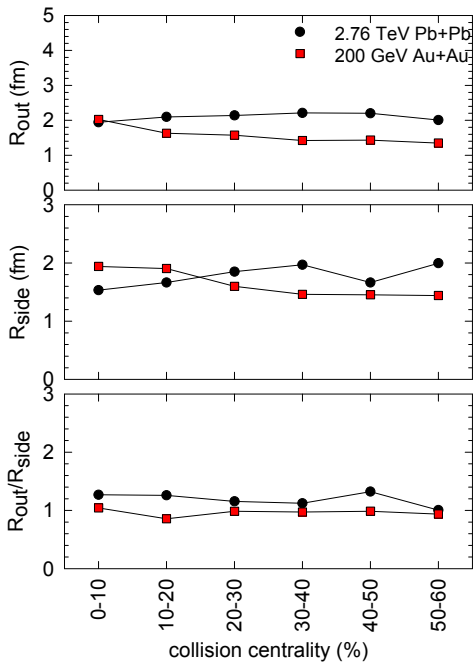}
}
\caption{(color online) The top and middle panels shows the simulated HBT radii $R_{out}$ and $R_{side}$ for direct photons  in $\sqrt{s}_{NN}$=2.76 TeV Pb+Pb and in $\sqrt{s}_{NN}$=200 GeV Au+Au collisions as a function of collision centrality. In the bottom   panel centrality dependence of the ratio $R_{out}/R_{side}$ is shown.
 }\label{F8}
\end{figure}

 From the correlation functions, we have also calculated the HBT radii, $R_{out}$ and $R_{side}$.
  Noting that the correlation functions depart from the Gaussian shape, we have defined the half width of the correlation function as the HBT radii, $R_{out}=\sqrt{2}/q^*_{out}$,  $R_{side}=\sqrt{2}/q^*_{side}$, where $q^*_{out}$ is determined from the condition,
 $2[C(q,K)-1]=0.5$ and similarly for $q^*_{side}$. Centrality dependence of HBT radii for direct photon emission is shown in Fig.\ref{F8}. The top panel shows the outward radius in  0-10\%-50-60\% Au+Au and Pb+Pb collisions.    Outward radius $R_{out}$, both in Pb+Pb collisions at LHC and in Au+Au collisions at RHIC, do not show significant centrality dependence. The behavior is unlike that observed for the charged particles.
HBT radii $R_{out}$ for the charged particles has been measured in $\sqrt{s}$=200 GeV Au+Au collisions  \cite{Adams:2004yc}\cite{Adler:2004rq}.   $R_{out}$ for the charged particles  shows significant centrality dependence, decreasing rapidly in peripheral collisions.  $R_{out}$ for photons however do show some energy dependence. With the exception of very central 0-10\% collisions
  $R_{out}$ is increased by 30-50\% from RHIC to LHC energy.  The result is not unexpected. $R_{out}$ measures emission size and duration of emission time. Emission time and duration supposed to be increased in LHC energy collision. However, we do note that while from RHIC to LHC, energy is increased by a factor of $\sim$14, $R_{out}$ in 10-20\%-50-60\% collisions is increased by $\sim$30-50\% only. Energy dependence is not large.

Centrality dependence of HBT sideward radius $R_{side}$ in $\sqrt{s}_{NN}$=2.76 TeV Pb+Pb
collisions and in $\sqrt{s}_{NN}$=200 GeV Au+Au collisions are shown in the middle panel of Fig.\ref{F8}.  
In 0-10\% -50-60\% collision, $R_{side}$ also do not show significant centrality dependence. Here again, the behavior is unlike that observed for the charged particles, where $R_{side}$  rapidly decreases in peripheral collisions. $R_{side}$ is a measure of transverse size of the emission zone, and is expected to be more in Pb+Pb collisions at LHC than in Au+Au collisions at RHIC (transverse expansion is more in Pb+Pb collisions than in Au+Au collisions). However, in the simulations,
in 0-10\% and 10-20\% collisions $R_{side}$ is more in Au+Au collisions than in Pb+Pb collisions. In more peripheral collisions however, as expected, $R_{side}$ is more in Pb+Pb collisions.  Why in central 0-10\% and 10-20\% collisions, direct photons are emitted from reduced sized fireball is not understood.

It is mentioned earlier that within some model assumptions,  the ratio $R_{out}/R_{side}$ is a measure of   the particle emission duration. The centrality dependence of the ratio in Pb+Pb collisions at LHC and in Au+Au collisions at RHIC is   shown in the bottom panel of Fig.\ref{F8}. 
The ratio is increased in Pb+Pb collisions at LHC. It appears that emission time for direct photons is increased in Pb+Pb collisions at LHC from that in Au+Au collisions at RHIC. However, the increase is not large, less than $\sim$50\%.  
One also notes that the while in Pb+Pb collisions, the ratio is marginally larger than unity, in Au+Au collisions, it is close to unity.

a 

The above HBT calculations are limited to photon momentum  $0<k_T<1 GeV$. Direct photon correlation and HBT radii do depend on the photon momentum. The dependence is studied in Fig.\ref{F9}, where simulation results for direct photon correlations, for $k_T$=0.5, 1.0, 1.5 and 2.0 GeV in 0-10\% Au+Au collisions at RHIC are shown. Both outside correlation show strong dependence on $k_T$, and side correlations, for 
$k_T$=0.5, 1.0, 1.5 and 2.0 GeV are stuid

\section{Summary}

To summarise, in ideal hydrodynamic model, we have simulated direct photon production in $\sqrt{s}_{NN}$=200 GeV Au+Au and in $\sqrt{s}_{NN}$=2.76 TeV Pb+Pb collisions. Ideal hydrodynamic model    simulations for direct photons reasonably well explain the PHENIX data on experimental   photon spectra in 0-10\%-40-50\% Au+Au collisions, upto $p_T\approx$3 GeV. At higher $p_T$ model produces less photons than in experiment. In more peripheral collisions, experimental data are explained only upto $p_T$=2 GeV. Ideal hydrodynamic model however produces less elliptic flow in Au+Au collisions than in experiment.
Ideal hydrodynamic model simulations also explain the ALICE measurements for the direct photon spectra in 0-40\% $\sqrt{s}_{NN}$=2.76 TeV Pb+Pb collisions.    In Pb+Pb collisions also the ideal hydrodynamic simulations produces less flow than in experiment at low $p_T$, $p_T \leq 3$GeV.  However, the discrepancy between model simulations and experiment is much less than that in Au+Au collisions at RHIC.  In higher $p_T$, model produces more flow than in experiment.

We have also simulated for HBT correlations for direct photons, 
  in Au+Au collisions at RHIC and Pb+Pb collisions at LHC. The correlation function was computed only for photons in the $p_T$ range 0-1 GeV. Outward and sideward radii, $R_{out}$ and $R_{side}$ was also extracted. Both in Au+Au and in Pb+Pb collision, in 0-10\% - 50-60\% collisions,   $R_{out}$ and $R_{side}$, for direct photons, do not show appreciable centrality dependence. The result is at variance with the centrality dependence of HBT radii for charged particles, where HBT radii decrease rapidly in peripheral collisions. Outward and sideward radius $R_{out}$ and $R_{side}$ for direct photons shows marginal energy dependence. From RHIC to LHC, while the collision energy 
is increased by a  factor of $\sim$14,  $R_{out}$, depending on the collision centrality, increases by 30-50\%. In   central 0-10\% and 10-20\% collisions, $R_{side}$ decreases by $\sim$10-20\% as the collision energy is increased, and in peripheral  20-30\%-50-60\% collisions, $R_{side}$ is increased by $\sim$30-40\%.
  The ratio $R_{out}/R_{side}$, which is considered as a measure of the emission time is  increased by $\sim$10-45\% from RHIC to LHC. Photons are emitted for longer duration in Pb+Pb collisions.
  
  In the present simulations, dissipative effects like shear viscosity are neglected.
However, from string theory motivated calculation  \cite{Policastro:2001yc}\cite{Kovtun:2003wp},  one understands that shear viscosity over entropy ratio has a lower bound, $\eta/s \geq 1/4\pi$. Charged particle production in Au+Au collisions at RHIC and in Pb+Pb collisions at LHC is consistent with small viscous fluid evolution. Viscous effect on photons production  was studied in  \cite{Dusling:2009bc}\cite{Chaudhuri:2011up}\cite{Dion:2011pp}. As it is for the charged particles, viscosity enhances $p_T$ production and decrease the elliptic flow.   As it is shown here, even in ideal hydrodynamic evolution, elliptic flow for photons is less than that in experiments. Agreement between experiment and simulations will further deteriorate if viscous effects are included. Indeed, it is mystery that  for direct photons, ideal hydrodynamic simulations produced much less elliptic flow  than in experiments. Evidently,
further investigations are required to understand photon production in relativistic heavy ion collisions.

\end{document}